\documentclass[prb,twocolumn,showpacs]{revtex4}
\usepackage{amsmath}
\usepackage{dcolumn}
\usepackage{graphicx}

\begin{document}

\title[Short title for running header]{Vanishing pseudogap around $(\pi,0)$ in an electron-doped high-$\mathrm{T_{c}}$ superconductor: a simple picture}
\author{Tao Li and Da-Wei Yao}
\affiliation{Department of Physics, Renmin
University of China, Beijing 100872, P.R.China}
\date{\today}

\begin{abstract}
Recent ARPES measurement on electron-doped cuprate $\mathrm{Pr}_{1.3-x}\mathrm{La}_{0.7}\mathrm{Ce}_{x}\mathrm{CuO}_{4}$ finds that the pseudogap along the boundary of the antiferromagnetic Brillouin zone(AFBZ) exhibits dramatic momentum dependence. In particular, the pseudogap vanishes in a finite region around the anti-nodal point, in which a single broadened peak emerges at the un-renormalized quasiparticle energy. Such an observation is argued to be inconsistent with the antiferromagnetic(AFM) band-folding picture, which predicts a constant pseudogap along the AFBZ boundary. On the other hand, it is claimed that the experimental results are consistent with the prediction of the cluster dynamical mean field theory(CDMFT) simulation on the Hubbard model, in which the pseudogap is interpreted as a s-wave splitting between the Hubbard bands and the in-gap states. Here we show that the observed momentum dependence of the pseudogap is indeed consistent with AFM band-folding picture, provided that we assume the existence of a strongly momentum dependent quasiparticle scattering rate. More specifically, we show that the quasiparticle scattering rate acts to reduce the spectral gap induced by AFM band-folding effect. The new quasiparticle poles corresponding to the AF-split bands can even be totally eliminated when the scattering rate exceeds the bare band folding gap, leaving the system with a single pole at the un-renormalized quasiparticle energy. We predict that the pseudogap should close in a square root fashion as we move toward $(\pi,0)$ along the AFBZ boundary. Our results illustrates again that the quasiparticle scattering rate can play a much more profound role than simply broadening the quasiparticle peak in the quasiparticle dynamics of strongly correlated electron systems. 
\end{abstract}

\maketitle
The origin of the pseudogap phenomena remains a major unresolved issue in the study of the high temperature superconductors\cite{Timusk,Lee,Kordyuk}. This phenomena has been interpreted either as a superconducting fluctuation effect, or, by many other researchers, as a precursor effect of some kind of competing order. While it is no doubt that the superconducting fluctuation effect does exist in high-$\mathrm{T_{c}}$ cuprates, it is by now quite clear that the superconducting fluctuation effect alone cannot be responsible for the whole story of the pseudogap phenomena. It is also generally believed that the strong correlation effect must be playing an essential role in the formation of the pseudogap phenomena.   

Among the various kind of competing order pictures proposed for the pseudogap phenomena, the antiferromagnetic spin fluctuation scattering picture is the most extensively studied. All cuprate superconductors are derived from doping antiferromagnetic insulating parent compounds. Extensive evidences have been accumulated through the years for the existence of strong antiferromagnetic spin fluctuation in the high-$\mathrm{T_{c}}$ superconductors\cite{Lee}. In particular, recent RIXS measurement shows that the spin-wave-like excitation at high energy is robust against doping even in the over-doped systems\cite{RIXS1,RIXS2,RIXS3,RIXS4,RIXS5,RIXS6}, with its dispersion and integrated intensity only slightly modified by doping. A theory of the pseudogap phenomena based on the scattering of the quasiparticle from the strong antiferromagnetic spin fluctuation has been studied more than two decades and is generally called the spin-Fermion model\cite{Chubukov}. This model can either be taken as a phenomenological model describing the interaction of the low energy quasiparticle with the long wave length antiferromagnetic spin fluctuation, or as a low energy effective theory of some underlying strongly correlated electron model in the renormalization group sense. The existence of robust local-moment-like degree of freedom at low energy can be interpreted as the main consequence of the strong correlation effect of the electron at higher energy. 

In the spin-Fermion model picture, the pseudogap phenomena is understood as a band folding effect induced by the quasiparticle scattering from the antiferromagnetic spin fluctuation. It should thus be the strongest along the boundary of the antiferromagnetic Brillouin zone(AFBZ), on which quasiparticles with momentum differ by $\mathrm{Q}=(\pi,\pi)$ have degenerate energies. The pseudogap induced by the AF scattering should thus be a constant on the AFBZ boundary. Such a simplification has been employed in a recent ARPES measurement on the electron-doped cuprate $\mathrm{Pr}_{1.3-x}\mathrm{La}_{0.7}\mathrm{Ce}_{x}\mathrm{CuO}_{4}$\cite{Hashimoto}, in which it is found that a momentum independent spectral gap can be clearly seen along most part of the AFBZ boundary. However, to one's surprise, the spectral gap is found to close abruptly in the anti-nodal region, in which the electron spectral function evolves into a single broadened peak\cite{edop1,edop2,edop3}. Such a counter-intuitive observation has been argued to be a strong evidence against the AFM band-folding picture by the authors of Ref.[\onlinecite{Hashimoto}], who alternatively relate the observed spectral gap to a s-wave splitting found in a cluster dynamical mean field theory(CDMFT) simulation of the Hubbard model. However, as will be detailed later in this paper, such an interpretation fails to account for some key features of the observed spectrum.     

In this work, we present an alternative and much simpler interpretation for the closing of the pseudogap in the anti-nodal region in the above electron-doped cuprate. In our picture, the observed pseudogap in $\mathrm{Pr}_{1.3-x}\mathrm{La}_{0.7}\mathrm{Ce}_{x}\mathrm{CuO}_{4}$ is still caused by the AFM band-folding effect. The dramatic momentum dependence of the observed pseudogap around $(\pi,0)$ is interpreted as a result of the competition between the AF band-folding effect and the quasiparticle incoherence effect. More specifically, we show that a finite quasiparticle scattering rate $\Gamma_{\mathrm{k}}$ not only broadens the quasiparticle peak, but also acts to reduce the size of the spectral gap induced by the AFM band-folding effect according to $\sqrt{\Delta_{AF}^{2}-\Gamma^{2}_{\mathrm{k}}}$. When the scattering rate exceeds the bare band-folding gap $\Delta_{AF}$, the spectral gap will be totally eliminated and a single broadened peak will emerge in the electron spectral function at the un-renormalized quasiparticle energy. Thus, the strange momentum dependence of the observed pseudogap is explained if we assume the existence of a large quasiparticle scattering rate in the anti-nodal region, which is indeed the case in all cuprates.

Before proceeding further, it is instructive to note the difference in the AF spin fluctuation in the electron- and the hole-doped cuprate superconductors. Strong particle-hole asymmetry in the phase diagram has long been noticed in the study of the high-$\mathrm{T_{c}}$ superconductivity\cite{Armitag}. In particular, it is found that the AF long range ordered region is much more extended in the electron-doped cuprates than in the hole-doped cuprates. Such an asymmetry can be naturally understood from the point of view of Fermi surface topology. More specifically, the Fermi surface in the electron-doped cuprates is more favorable(better nested) for AF ordering than that of the hole-doped cuprates. As a result, the AF spin correlation exists mainly in the form of long range and static order in the electron-doped cuprates but more like a short-ranged and dynamical fluctuation in the hole-doped cuprates. Such a difference should result in different pseudogap behavior in both types of high-$\mathrm{T_{c}}$ cuprates.

To compare with previous study more closely, we adopt exactly the same quasiparticle model as was used in Ref.[\onlinecite{Hashimoto}], which is given by
\begin{equation}
H_{MF}=\sum_{\mathrm{k},\sigma}\xi_{\mathrm{k}}c^{\dagger}_{\mathrm{k},\sigma}c_{\mathrm{k},\sigma}+\Delta_{AF}\sum_{\mathrm{k},\sigma}\sigma c^{\dagger}_{\mathrm{k+Q},\sigma}c_{\mathrm{k},\sigma}.\nonumber
\end{equation}
Here $\xi_{\mathrm{k}}=-2t(\cos k_{x}+\cos k_{y})-4t'\cos k_{x}\cos k_{y}-2t''(\cos2k_{x}+\cos2k_{y})+\mu$ describes the bare quasiparticle band structure. According to Ref.[\onlinecite{Hashimoto}], a best fit to the spectra of $\mathrm{Pr}_{1.3-x}\mathrm{La}_{0.7}\mathrm{Ce}_{x}\mathrm{CuO}_{4}$ is achieved for $t=0.25 eV$, $t'/t=-0.15$, $t''/t'=-0.5$ and $\mu/t=-0.02$. $\Delta_{AF}$ denotes the magnitude of the AF long range order in the system and $\mathrm{Q}=(\pi,\pi)$ is the AF wave vector. As was found in Ref.[\onlinecite{Hashimoto}], a best fit of the AF splitting in the electron spectrum is achieved for  $\Delta_{AF}=0.11 \ eV$. 

We note that in principle one can also assume the system to have a finite spin correlation length of $\xi$, rather than being AF long range ordered. However, it is straightforward to show that a rather large $\xi$ is needed to account for the well defined AF band-folding gap in the nodal direction. More specifically, we expect that $\frac{2\pi v_{F}}{\xi}\ll\Delta_{AF}$, otherwise the AFM band-folding gap would be strongly suppressed by the detuning effect of the quasiparticle dispersion in the nodal direction. Here $v_{F}$ denotes the Fermi velocity in the nodal direction, which can be estimated as $v_{F}\approx4\sqrt{2}t=1.44 \ eV$(here we use the Fermi velocity at $\mathrm{k}=(\pi/2,\pi/2)$ to approximate the Fermi velocity along the nodal direction). Inserting this estimation into $\frac{2\pi v_{F}}{\xi}\ll\Delta_{AF}$, one find that the spin correlation length should be much larger than 20 lattice constants to produced the observed spectral gap in the nodal region. It is thus quite consistent to assume a true AF long range order in the system.

According to the AFM band-folding picture, the quasiparticle band along the AFBZ boundary should be split into two branches with the dispersion $\xi_{\mathrm{k}}\pm\Delta_{AF}$. Such a description fails to account for the dramatic momentum dependence of the pseudogap in the anti-nodal region. To solve this problem, we assume the quasiparticle to have a strongly momentum dependent scattering rate that peaks in the anti-nodal region. As we will show below, the quasiparticle scattering rate acts to reduce the pseudogap induced by the AFM band-folding effect. For convenience, we assume that the quasiparticle scattering rate to take the form of 
\begin{equation}
\Gamma_{\mathrm{k}}=\Gamma_{0}+\Gamma_{1}\sum_{i}\exp(-\eta(\mathrm{k-Q}_{i})^{2}).\nonumber
\end{equation}
Here $\mathrm{Q}_{i}=(\pm\pi,0)$ or $(0,\pm\pi)$ are the four anti-nodal points in the Brillouin zone, $\eta$ is a parameter introduced to define the size of the anti-nodal region, $\Gamma_{0}$ is a constant background. The scattering rate so defined reaches its maximum at the four anti-nodal points along the AFBZ boundary. When $\Gamma_{\mathrm{k}}>\Delta_{AF}$, the splitting caused by the AFM band-folding effect will be wiped out and we would expect a single broadened peak to emerge in the electron spectral function.

More precisely, the electron Green's function in the absence of the antiferromagnetic long range order is assumed to be given by
\begin{eqnarray}
G^{(0)}(\mathrm{k},\omega)=\frac{1}{\omega+i\Gamma_{\mathrm{k}}-\xi_{\mathrm{k}}}.\nonumber
\end{eqnarray}  
The self-energy corresponding to the AFM scattering is given by
\begin{eqnarray}
\Sigma(\mathrm{k},\omega)=\Delta^{2}_{AF}G^{(0)}(\mathrm{k+Q},\omega).\nonumber
\end{eqnarray}
Thus the Green's function of the renormalized quasiparticle is given by
\begin{eqnarray}
G(\mathrm{k},\omega)=\frac{1}{\omega+i\Gamma_{\mathrm{k}}-\xi_{\mathrm{k}}-\frac{\Delta^{2}_{AF}}{\omega+i\Gamma_{\mathrm{k+Q}}-\xi_{\mathrm{k+Q}}}}.\nonumber
\end{eqnarray}

On the AFBZ boundary, we have $G^{(0)}_{\mathrm{k},\omega}=G^{(0)}(\mathrm{k+Q},\omega)$, so that we have 
\begin{eqnarray}
G(\mathrm{k},\omega)=\frac{1}{\omega+i\Gamma_{\mathrm{k}}-\xi_{\mathrm{k}}-\frac{\Delta^{2}_{AF}}{\omega+i\Gamma_{\mathrm{k}}-\xi_{\mathrm{k}}}}.\nonumber
\end{eqnarray}
The electronic spectral function is given by the imaginary part of the Green's function and takes the form of
\begin{eqnarray}
A(\mathrm{k},\omega)&=&-2\mathrm{Im} G(\mathrm{k},\omega)\nonumber \\
&=& \frac{2\Gamma_{\mathrm{k}}[1+R(\mathrm{k},\omega)]}{(\omega-\xi_{\mathrm{k}})^{2}[1-R(\mathrm{k},\omega)]^{2}+\Gamma^{2}_{\mathrm{k}}[1+R(\mathrm{k},\omega)]^{2}},\nonumber
\end{eqnarray}
in which $R(\mathrm{k},\omega)=\frac{\Delta^{2}_{AF}}{(\omega-\xi_{\mathrm{k}})^{2}+\Gamma^{2}_{\mathrm{k}}}$.

From this expression, we see that the new poles induced by the antiferromagnetic band-folding effect is determined by the equation $R(\mathrm{k},\omega)=1$, whose solution is given by $\omega_{\pm}=\xi_{\mathrm{k}}\pm\sqrt{\Delta^{2}_{AF}-\Gamma^{2}_{\mathrm{k}}}$. Thus a finite scattering rate $\Gamma_{\mathrm{k}}$ not only broadens the quasiparticle peak in the electron spectrum, but also acts to reduce the spectral gap induced by AFM band-folding effect. In particular, when $\Gamma_{\mathrm{k}}>\Delta_{AF}$, these new poles will disappear, leaving the system with an un-renormalized pole at $\xi_{\mathrm{k}}$.  

To be more quantitative, we plot in Figure 1 the electronic spectral function along the AFBZ boundary for $\Gamma_{1}=0$ and $\Gamma_{1}=2\Delta_{AF}$. Here we set $\Gamma_{0}=0.03 \ eV$ and $\eta=2$ in both cases. When $\Gamma_{1}=0$, we find two split spectral peaks along the whole AFBZ boundary. When $\Gamma_{1}=2\Delta_{AF}$, the spectral gap is found to close in a finite region around the anti-nodal point, but remains almost momentum independent outside the anti-nodal region, just as what is observed in ARPES experiment. We believe this mechanism explains the strange momentum dependence of the pseudogap observed in $\mathrm{Pr}_{1.3-x}\mathrm{La}_{0.7}\mathrm{Ce}_{x}\mathrm{CuO}_{4}$.
\begin{figure}
\includegraphics[width=10cm]{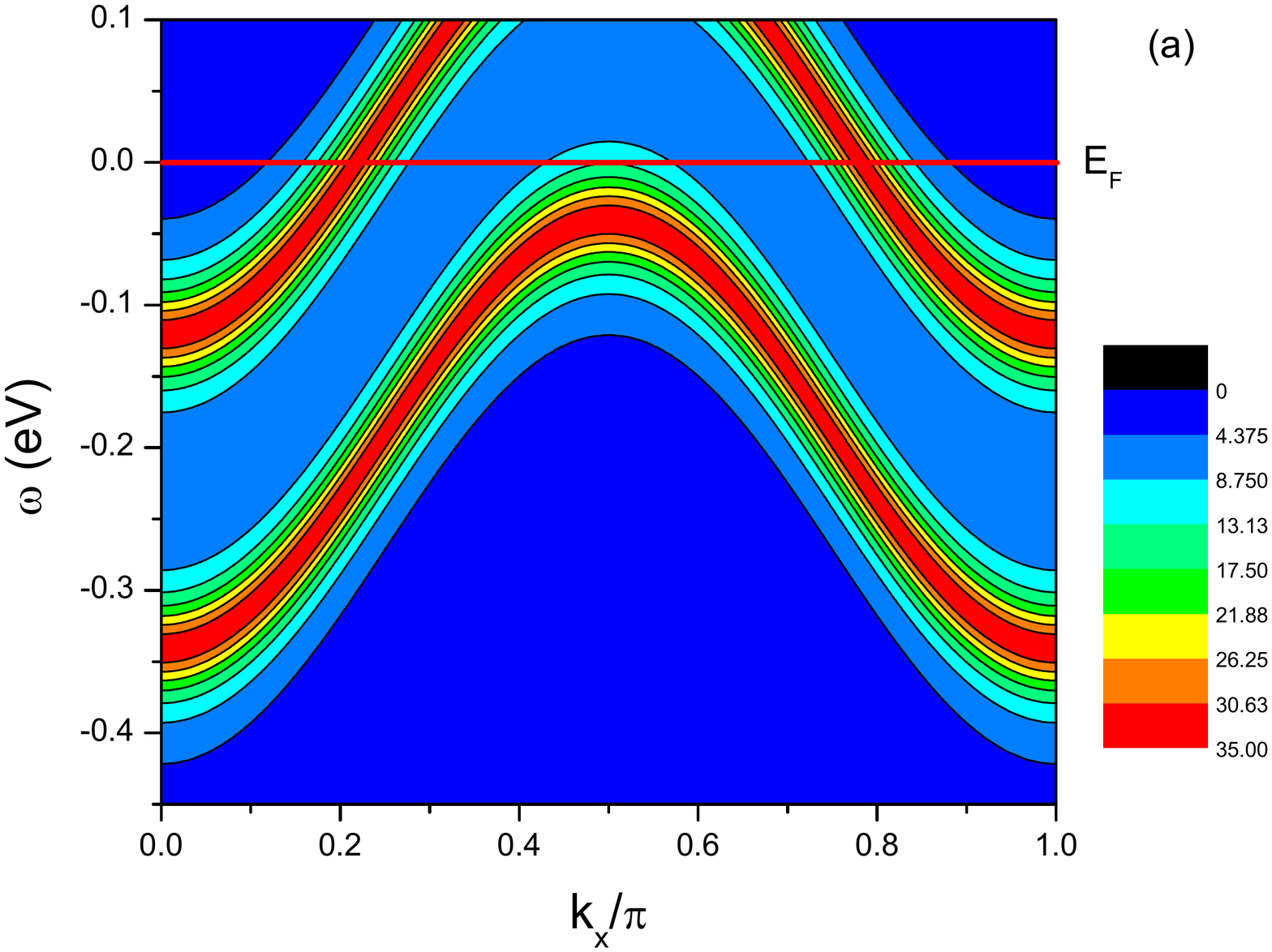}
\includegraphics[width=10cm]{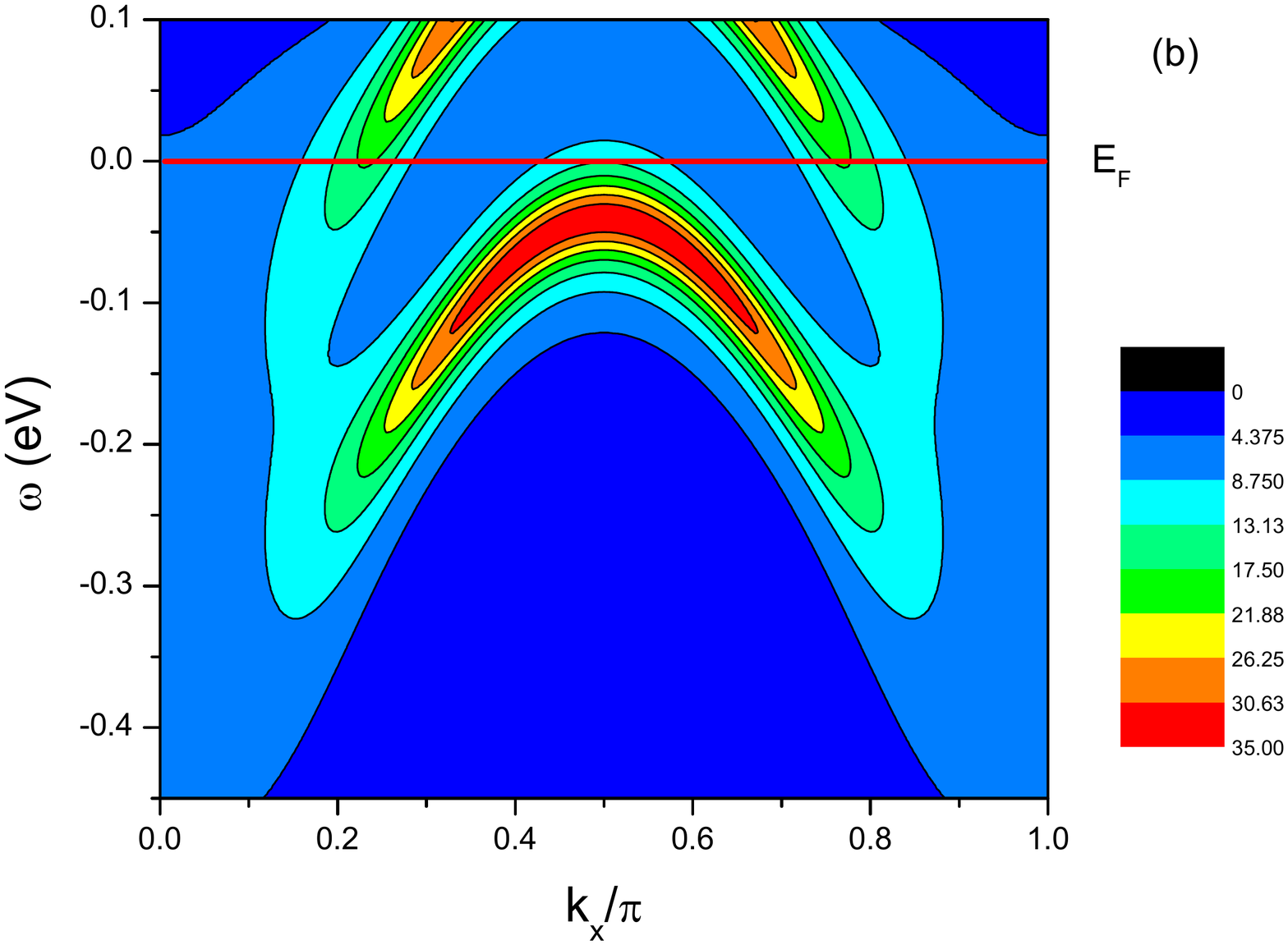}
\caption{\label{fig1}
(Color on-line) Intensity map of the electron spectral function along the AFBZ boundary. (a)The spectral function for $\Gamma_{1}=0$, (b)The spectral function for $\Gamma_{1}=2\Delta_{AF}=0.22 eV$. The red horizontal line indicate the position of the Fermi level.}
\end{figure}

To see more clearly the dispersion of the renormalized quasiparticles, we plot in Figure 2 the peak position of the electron spectral function as a function of $k_{x}$ along the AFBZ boundary. One finds that the splitting induced by antiferromagnetic band-folding effect decreases continuously as one approaches the anti-nodal point. When $\mathrm{k}$ is sufficiently close to the anti-nodal point so that the scattering rate $\Gamma_{\mathrm{k}}$ is greater than $\Delta_{AF}$, the two spectral peaks merge into a single broaden peak at the un-renormalized quasiparticle energy $\xi_{\mathrm{k}}$, just as what is observed in the ARPES experiment.
\begin{figure}
\includegraphics[width=9cm]{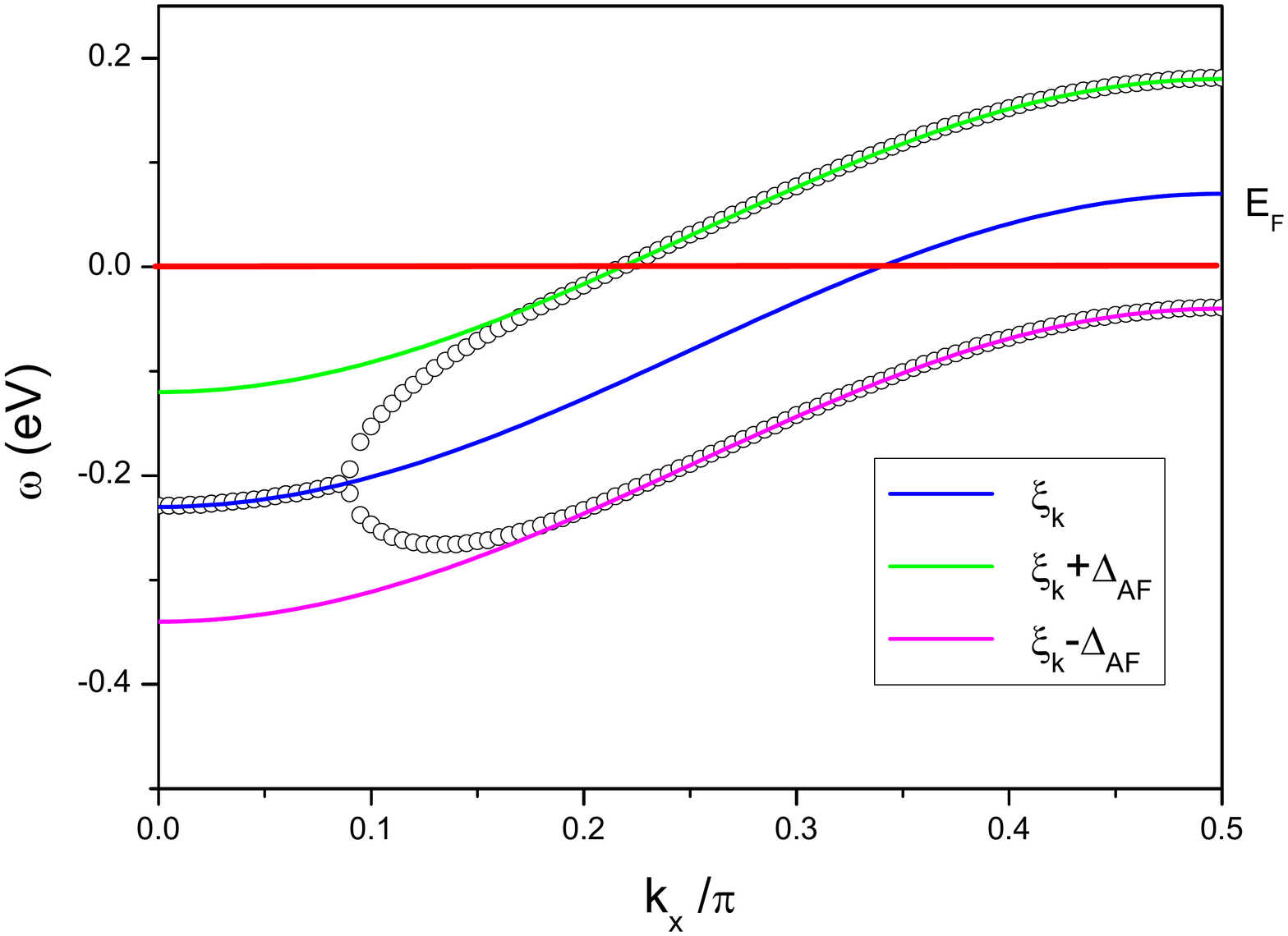}
\caption{\label{fig1}
(Color on-line) Peak positions(shown as open circles) in the electron spectral function as a function of $k_{x}$ along the AFBZ boundary. The blue line denotes the bare dispersion $\xi_{\mathrm{k}}$, the green and pink lines are the quasiparticle dispersion in the presence of the antiferromagnetic order. Here we set $\Gamma_{1}=2\Delta_{AF}=0.22 eV$, $\Gamma_{0}=0.03 eV$.}
\end{figure}

From the above results, we see that the dramatic momentum dependence of the pseudogap observed in the electron doped cuprate $\mathrm{Pr}_{1.3-x}\mathrm{La}_{0.7}\mathrm{Ce}_{x}\mathrm{CuO}_{4}$ can indeed be understood in the antiferromagnetic band-folding picture, provided that we assume the existence of a strongly momentum dependent quasiparticle scattering rate along the AFBZ boundary. Our analysis shows that to develop a well defined spectral gap in the AFM band-folding picture, the quasiparticle scattering rate $\Gamma_{\mathrm{k}}$ and the detuning caused by quasiparticle dispersion, namely, $\frac{2\pi v_{{F}}}{\xi}$, should be both smaller than the bare folding gap $\Delta_{AF}$. In particular, the observation of the clear spectral gap in the nodal direction in the electron doped cuprate $\mathrm{Pr}_{1.3-x}\mathrm{La}_{0.7}\mathrm{Ce}_{x}\mathrm{CuO}_{4}$ indicates that the spin correlation in this system should be rather long-ranged, if not truely long range ordered.

The existence of the strongly momentum dependent quasiparticle scattering rate is a phenomenological assumption in our study. However, such an assumption is not only an observed fact for the cuprate superconductors, but is also very natural from the point of view of microscopic models. More specifically, since the band effective mass diverges at the anti-nodal points, it is natural to expect the electron correlation effect to be stronger in the anti-nodal region. We think such a strongly momentum dependent scattering rate comes most likely from the scattering by short-ranged and dynamical spin fluctuation, which exists universally in the high-$\mathrm{T_{c}}$ cuprate superconductors. 

Here we note that the detuning effect induced by the quasiparticle dispersion and the diffusion in the energy of the dynamical spin fluctuation is playing a similar role as a finite quasiparticle scattering rate in reducing the AFM band-folding gap. Such detuning effect is much more significant in the hole-doped cuprates, in which the spin fluctuation is more short-ranged and more dynamical in nature than that in the electron-doped cuprates. We think this difference is responsible for the absence of AFM band-folding gap in the nodal region in the hole-doped cuprates. We also believe that the AFM band-folding effect alone is not enough to account for the pseudogap phenomena in the hole-doped cuprates\cite{Li}. This implies that the pseudogap observed in the hole-doped cuprates may have different origin from the AFM band-folding gap in the electron-doped cuprates.

In Ref.[\onlinecite{Hashimoto}], the authors claimed that a CDMFT simulation on the Hubbard model can produce a spectral signature that is similar to the pseudogap observed in $\mathrm{Pr}_{1.3-x}\mathrm{La}_{0.7}\mathrm{Ce}_{x}\mathrm{CuO}_{4}$. In the CDMFT, the pseudogap is attributed to the splitting between the lower/upper Hubbard band and the in-gap state in the hole/electron-doped cuprates, which exhibits a s-wave character on the Fermi surface with moderate momentum dependence. However, such an interpretation suffers from the following problems. First, the CDMFT predicts that the spectral intensity reaches its maximum in the anti-nodal region along the AFBZ boundary, while the observed spectral intensity is the weakest there. Second, while the CDMFT predicts a moderate momentum dependence of the pseudogap along the AFBZ boundary, the observed momentum dependence is much more dramatic. More specifically, the observed pseudogap seems to vanish abruptly in the anti-nodal region and is almost momentum independent outside the anti-nodal region. Our theory provides a consistent explanation for both of these two characteristics. Our theory also predicts that the pseudogap will close in a square root fashion in the anti-nodal region. This should be checked in future experimental studies. 

The results presented in this work indicate that the quasiparticle scattering rate can play a much more profound role than simply broadening the quasiparticle peak in the quasiparticle dynamics of strongly correlated electron systems. We note that in a series of recent works, the same physics has been invoked to predict new state of matters in Dirac Fermion systems, topological insulators and Kondo systems\cite{Kozii,Zhou,Papaj,Fu,Qi}. In particular, it is found that a strongly momentum dependent quasiparticle scaterring rate may be responsible for the emergence of Fermi arc in the bulk dispersion of the correlated electron systems. It is interesting to see if such a mechanism has any relevance to the Fermi arc phenomena in the high-T$_{c}$ cuprates. 

In conclusion, we proposed an alternative explanation for the vanishing of the pseudogap around the anti-nodal point in the electron-doped cuprate $\mathrm{Pr}_{1.3-x}\mathrm{La}_{0.7}\mathrm{Ce}_{x}\mathrm{CuO}_{4}$. According to our picture, the pseudogap in such an electron-doped cuprates is still induced by the AFM band-folding effect. We show that the momentum dependence of the pseudogap is actually a manifestation of the competition between the AFM band-folding effect and the quasiparticle incoherence effect. In particular, the pseudogap in $\mathrm{Pr}_{1.3-x}\mathrm{La}_{0.7}\mathrm{Ce}_{x}\mathrm{CuO}_{4}$ is wiped out in the anti-nodal region because of the large quasiparticle scattering rate in this region.


\begin{thebibliography}{99}
\bibitem{Qi}We are in debt to Yang Qi for drawing our attention to the related works [\onlinecite{Kozii,Zhou,Papaj,Fu}] on the effect of sacttering rate in the quasiparticle dynamics of strongly correlated electron systems.
\end{thebibliography}
\end{document}